\documentclass[aps,preprint,showpacs,floatfix]{revtex4}
\usepackage{epsfig}
\usepackage{amsmath,amsfonts,bm}
\begin{document}

\count255=\time\divide\count255 by 60 \xdef\hourmin{\number\count255}
  \multiply\count255 by-60\advance\count255 by\time
 \xdef\hourmin{\hourmin:\ifnum\count255<10 0\fi\the\count255}

\newcommand\<{\langle}
\renewcommand\>{\rangle}
\renewcommand\d{\partial}
\newcommand\LambdaQCD{\Lambda_{\textrm{QCD}}}
\newcommand\tr{\mathop{\mathrm{Tr}}}
\newcommand\+{\dagger}
\newcommand\g{g_5}

\newcommand{\xbf}[1]{\mbox{\boldmath $ #1 $}}

\title{Pion Form Factors in Holographic QCD}

\author{Herry J. Kwee}
\email{Herry.Kwee@asu.edu}

\author{Richard F. Lebed}
\email{Richard.Lebed@asu.edu}

\affiliation{Department of Physics, Arizona State University, Tempe,
AZ 85287-1504}

\date{December 2007}

\begin{abstract}
Using a holographic dual model of QCD, we compute the pion
electromagnetic form factor $F_\pi (Q^2)$ in the spacelike momentum
transfer region, as well as pion couplings to vector mesons
$g_{\rho^{(n)} \pi \pi}$.  Spontaneous and explicit chiral symmetry
breaking are intrinsic features of this particular holographic model.
We consider variants with both ``hard-wall'' and ``soft-wall''
infrared cutoffs, and find that the $F_\pi (Q^2)$ data tend to lie
closer to the hard-wall model predictions, although both are too
shallow for large $Q^2$.  By allowing the parameters of the soft-wall
model (originally fixed by observables such as $m_\rho$) to vary, one
finds fits that tend to agree better with $F_\pi (Q^2)$.  We also
compute the pion charge radius $\langle r_\pi^2 \rangle$ for a variety
of parameter choices, and use the values of $f^{(n)}_\rho$,
$g_{\rho^{(n)} \pi \pi}$ and $m^{(n)}_\rho$ to observe the saturation
of $F_\pi (0)$ by $\rho$ poles.
\end{abstract}


\maketitle

\section{Introduction} \label{intro}

Quantum chromodynamics, now in its fourth decade, has been known since
its inception to stubbornly resist direct analytical solutions in its
strong coupling regime.  The modern period has seen numerous
approaches developed to tackle this problem, sometimes by providing a
simplified picture of strong interactions and its states (e.g., quark
potential models, chiral Lagrangians, quenched lattice calculations),
or by working in energy or mass regimes where a key parameter may be
assumed small (e.g., asymptotic freedom regime calculations,
operator-product expansions, heavy quark effective theory), or by
studying features of quantum field theories that incorporate
distinctly nonperturbative behavior (e.g., solitons, instantons).

Perhaps the most interesting techniques are ones that stimulate
advances in studies of QCD-like theories by making them {\em more\/}
complicated by introducing additional degrees of freedom. In this
category one includes the $1/N_c$ expansion and, more recently, the
gravity/gauge correspondence known by the names of the
anti-de~Sitter/conformal field theory (AdS/CFT) correspondence or the
holographic dual approach~\cite{AdSCFT}.  This proposed duality
between strongly-coupled Yang-Mills theories and weakly-coupled
gravity is exceptionally appealing because it implies a fundamental
connection between gauge and string theories.

The original example of this duality is given by ${\cal N} = 4$
super-Yang Mills theory, which is conformal and therefore lacks
asymptotic $S$-matrix particle states.  While this hardly seems like
an auspicious starting point for modeling QCD-like theories and their
rich hadronic spectra, a number of the original dual theories
nonetheless possess such useful QCD-like properties as chiral symmetry
breaking and confinement.  Moreover, approximate conformal symmetry is
a well-known property of QCD in the deep ultraviolet (UV) limit.

In the holographic approach one begins with the 5-dimensional AdS
metric,
\begin{equation}
ds^2 = g^{\vphantom\dagger}_{MN} \, dx^M dx^N = \frac{1}{z^2}
(\eta_{\mu \nu} dx^\mu dx^\nu - dz^2) \,
, \label{metric}
\end{equation}
where $\eta_{\mu \nu} \! = \! \rm{diag} (+,-,-,-)$ is distinguished
from the full nontrivial 5D metric $g^{\vphantom\dagger}_{MN}$
obtained from Eq.~(\ref{metric}).  The $z$ (Liouville or ``bulk'')
coordinate corresponds to an inverse energy scale ($Q \! \sim \!
1/z$), in that the UV limit of QCD is represented by fields living on
the AdS boundary $z \! = \! 0$ (or, allowing for a UV cutoff, a small
finite value $z \! = \! \epsilon$, a location called the ``UV
brane'').  The gauge/gravity correspondence then states that every CFT
operator ${\cal O}(x)$ is associated with a bulk field $\Psi (x,z)$
uniquely determined by its value $\Psi (x, \epsilon)$ on the UV brane.
The conformal symmetry is broken, introducing thereby a mass scale, by
limiting the ability of the fields $\Psi (x,z)$ to penetrate deeply
into the bulk, which corresponds to constraining the infrared (IR)
behavior; this may be accomplished, for example, by imposing a hard
cutoff and appropriate boundary conditions on $\Psi (x,z)$ at a value
$z \! = \! z_0$ (called the ``IR brane'')~\cite{PS} or by introducing
a soft wall with an exponential decrease $\sim e^{-\kappa^2 z^2}$ in
the action for large $z$~\cite{KKSS}.  The dimensionful parameters
$z_0^{-1}$ or $\kappa$ consequently serve the role of $\Lambda_{\rm
QCD}$.  The Kaluza-Klein modes of the field $\Psi (x,z)$ then
represent hadronic states of the same quantum numbers, producing
towers of hadrons analogous to those arising in the original hadronic
flux-tube string theories of the 1970's, which in turn are in close
kinship with the original Regge theories of hadronic excitations.

Fully exploiting the gauge/gravity correspondence to produce a model
for real strong interaction physics---a method called ``holographic
QCD'' or ``AdS/QCD''---may be attempted either through a top-down
approach starting with a particular string theory and choosing a
background that (as mentioned above) naturally produces QCD-like
properties, or a bottom-up approach starting with real QCD properties
and using them to obtain constraints on viable dual gravity theories.
In this paper we adopt the latter viewpoint.  Work along these lines
has become very popular in the past couple of years; especially well
represented and close to this work in spirit are studies of hadronic
spectra~\cite{EKSS,KKSS,BoschiFilho:2002vd,de
Teramond:2005su,Evans:2006ea,Hong:2006ta,Colangelo:2007pt,
Forkel:2007cm}, the couplings of hadrons in the presence of chiral
symmetry breaking~\cite{EKSS,Da
Rold:2005zs,Hirn:2005nr,Ghoroku:2005vt,HuangZuo}, and hadronic form
factors~\cite{Hong:2004sa,GR1,GR2,Radyushkin:2006iz,BdT}.  In this
paper we are specifically interested in employing the formalism
introduced in Ref.~\cite{EKSS}, in which chiral symmetry breaking is
included directly in the Lagrangian, to obtain specific information on
the form factor of the pion in this variant of AdS/QCD.

According to the holographic correspondence, the global QCD symmetry
of isospin associated with the two light quark flavors is promoted to
a gauged SU(2) symmetry respected by the bulk fields, and the problem
reduces to one of constructing a 5D action containing all the fields
of interest with the appropriate quantum numbers and Lorentz
structures.  Varying the action with respect to the fields of interest
generates their wave equations, which are solved as eigenvalue
equations subject to appropriate boundary conditions to obtain the
modes (mesons with particular masses and $z$-dependent wave
functions).  Meson decay constants appear as values of the modes at
the UV boundary (associated, as usual, with wave functions at small
distance scales) and form factors appear as convolution integrals over
$z$ of wave functions and currents.

The behavior of wave functions and form factors of vector mesons (the
$\rho$ and its partners) has been examined using the constructions of
Ref.~\cite{EKSS} using both hard-wall~\cite{GR1} and
soft-wall~\cite{GR2} IR boundary conditions.  The former has the
advantage of simplicity but produces the unphysical Regge trajectory
$M_n^2 \sim n^2$, while the latter produces the more
phenomenologically realistic behavior $M_n^2 \sim n^1$~\cite{KKSS}.  A
number of interesting results follow from the analysis of
Refs.~\cite{GR1,GR2}, including a distinctive pattern of vector meson
dominance by the lowest-mass states and predictions of the $\rho$
charge radius.  However, the calculations of Refs.~\cite{GR1,GR2}
require only one dimensionful parameter, $z_0$ or $\kappa$,
respectively, and their eigenvalue equations for the vector mesons
admit closed-form analytic solutions in terms of known functions
(Bessel functions and Laguerre polynomials).  The pion sector as
described in Ref.~\cite{EKSS} also requires dimensionful parameters
associated with both spontaneous ($\sigma$) and explicit ($m_q$) chiral
symmetry breaking, and the resultant wave equations cannot be solved in
closed form for arbitrary values of the three dimensionful parameters.  It
is the purpose of this paper to examine both the analytic limiting cases
of the pion interpolating field and numerical solutions for pion
couplings and form factor, in both the hard- and soft-wall cases.

In our numerical simulations we find that the data for the pion form
factor $F_\pi (Q^2)$ lies closer to the hard-wall than the soft-wall
model results.  The three parameters of the hard-wall model were
originally fit to $m_\rho$, $m_\pi$, and $f_\pi$, and we repeat the
exercise for the parameters of the corresponding soft-wall model.  One
may adjust the parameters of either model to obtain a better fit to
$F_\pi (Q^2)$, but at the cost of a poorer fit to at least one of
these three observables.
The $F_\pi (Q^2)$ data suggests an optimal model that incorporates
features of both, but is closer to the hard-wall model.

A calculation~\cite{GR3} using the same formalism of Ref.~\cite{EKSS}
considers $F_\pi(Q^2)$ in the chiral limit ({\it i.e.}, sets $m_q \! =
\! 0$) and focuses primarily on analytical behavior.  We agree with
their finding that the original hard-wall model overshoots $F_\pi
(Q^2)$ data for large $Q^2$ (but not with their choice of
normalization for parameters such as $f_\pi$).

This paper is organized as follows: In Sec.~\ref{formalism} we recount
and extend the formalism of Ref.~\cite{EKSS} relevant to our
calculations and exhibit the analytically soluble limits to the
equations of motion for the field containing the pion modes.
Section~\ref{newFF} gives expressions for the pion form factor and
couplings in terms of AdS/QCD mode wave functions.
Section~\ref{results} presents the results of a number of numerical
simulations of the pion electromagnetic form factor and couplings to
vector mesons, and Sec.~\ref{concl} summarizes our results and
concludes.

\section{Formalism} \label{formalism}

The full 5-dimensional action~\cite{EKSS} used in this work reads
\begin{equation}\label{5DL}
S = \int\! d^{\, 5} \! x \, e^{-\Phi} \! \sqrt{g}\, \tr \left\{ |DX|^2
+ 3 |X|^2 - \frac1{4\g^2} (F_L^2 + F_R^2) \right\} \, ,
\end{equation}
where $g \! \equiv \! | \det g^{\vphantom\dagger}_{MN} |$ is obtained
from the metric in Eq.~(\ref{metric}), and $e^{-\Phi}$ represents a
background dilaton coupling, with $\Phi (z) \! = \! 0$ in the
hard-wall case and $\kappa^2 z^2$ in the soft-wall case.  The
Lagrangian within the braces of Eq.~(\ref{5DL}) is written in terms of
a scalar field $X$ and chiral gauge fields $A_{L,R}$ that enter
through $D^M X \equiv \d^M \! X \! - iA_L^M X + iX A_R^M$, $A_{L,R}^M
\equiv A_{L,R}^{M \, a} \, t^a$ with $t^a$ being the generators of the
gauged isospin symmetry, and $F_{L,R}^{MN} \! \equiv \d^M \! A_{L,R}^N
\! - \d^N \! A_{L,R}^M \! - i[A_{L,R}^M , A_{L,R}^N]$.  The chiral
gauge fields $A^{\mu \, a}_{L,R}$ are the holographic partners of the
QCD operators $\bar q^{\vphantom\dagger}_{L,R} \, \gamma^\mu t^a
q^{\vphantom\dagger}_{L,R}$, while $X^{\alpha \beta}$ [more precisely
$(2/z)X^{\alpha \beta}$] is associated with $\bar q^\alpha_R \,
q^\beta_L$, and therefore incorporates all chiral symmetry-breaking
behavior.  In the holographic dictionary of Ref.~\cite{EKSS}, the
vacuum expectation value $X_0$ of $X$ (exact for the hard-wall model)
is given by
\begin{equation}
  X_0(z) = \frac{1}{2}M
       z + \frac{1}{2} \Sigma \,  z^3 \, , \label{eq:vev}
\end{equation}
where $M \! = \! m_q \openone$ and $\Sigma \! = \! \sigma
\openone$ represent explicit and spontaneous chiral symmetry breaking,
respectively, and arise in the holographic recipe through the
normalizable and nonnormalizable solutions for the bulk field $X$.
Strictly speaking, Eq.~(\ref{eq:vev}) holds for the soft-wall model
only for small values of $z$~\cite{KKSS}; for large $z$ the
corresponding $X_0$ should approach a constant, but the $e^{-\kappa^2
z^2}$ background minimizes this distinction.  We perform subsequent
calculations using a background field that formally satisfies both
limiting forms in Ref.~\cite{KL2}; however, as shown there, the best
fits do not improve upon the naive soft-wall fits presented below.
The pion field $\pi^a$ appearing through $X \!  = \! X_0 \exp (2i
\pi^a t^a)$ is dimensionless and related to the canonically-normalized
pion field $\tilde \pi^a$ of chiral Lagrangians via $\pi^a \! = \!
\tilde \pi^a / f_\pi$, with $f_\pi \! = \! 93$~MeV.

One now forms the polar and axial gauge fields $V,A$: $V^M \! \equiv
\! \frac 1 2 (A_L^M \! + A_R^M)$ and $A^M \! \equiv \! \frac 1 2
(A_L^M \! - A_R^M)$, in terms of which $D^M X = \d^M \! X -
i [V^M \! , X] - i \{A^M , X \}$, $F_V^{MN} \equiv
\d^M V^N \! - \d^N V^M \! - i \left( [V^M \! , V^N] +
[A^M \! , A^N] \right)$, $F_A^{MN} \equiv \d^M \! A^N \! - \d^N \! A^M
\! - i \left( [V^M \! , A^N] + [A^M \! , V^N] \right)$, and
\begin{equation}\label{5DL_V_A}
S = \int\! d^{\, 5} \! x \, e^{-\Phi} \! \sqrt{g} \, \tr \left\{
|DX|^2 + 3 |X|^2 - \frac{1}{2\g^2} (F_V^2 + F_A^2) \right\} \, .
\end{equation}
This action represents the only terms quadratic in fields and
containing two derivatives or less, and is therefore sufficient to
obtain the free-field equations of motion for $V_M$, $A_M$, and $\pi$
[generically $\Psi (x,z)$].  It is convenient to work in an axial-like
gauge, $V_z (x,z) \! = \! 0$, $A_z (x,z) \! = \! 0$; the associated
sources may then be expressed as divergences over only the usual four
spacetime dimensions, $\d^\mu V_\mu \! = \! 0$ (since isospin is
conserved) and $\d^\mu \! A_\mu$.  $A_\mu$ is further decomposed into
a transverse (divergenceless) piece $A_{\mu \, \perp}$ and a
longitudinal piece $\varphi$: $A_\mu \! = \! A_{\mu \, \perp} +
\d_\mu \varphi$.  The fields are chosen to satisfy Neumann conditions
$\d_z \Psi(x,z) \! = \! 0$ at $z \! = \! z_0$ in the hard-wall case
and to give a vanishing contribution from the $z \! \to \! \infty$ limit
in the soft-wall case.

Solving for the equations of motion of the fields $\Psi (q,z)$ Fourier
transformed with respect to the 4D coordinates $x$ yields
\newcommand{\AT}{A^a_\mu}
\begin{equation}\label{eqVAdS}
\d_z\left(\frac{e^{-\Phi}}{z} \, \d_z V_\mu^a \right) 
 + \frac{q^2 e^{-\Phi}}z V_\mu^a = 0 \, ,
\end{equation}
\begin{equation}
  \left[ \d_z\left(\frac{e^{-\Phi}}{z} \, \d_z \AT \right) +
\frac{q^2 e^{-\Phi}}{z}
  \AT - \frac{\g^2 \, v(z)^2 e^{-\Phi}}{z^3} \AT\right]_\perp =0 \, ,
\label{AT}
\end{equation}
\begin{equation}
  \d_z\left(\frac{e^{-\Phi}}{z} \, \d_z \varphi^a \right) 
+\frac{\g^2 \, v(z)^2 e^{-\Phi}}{z^3} (\pi^a-\varphi^a) = 0 \, ,
\label{AL}
\end{equation}
\begin{equation}
  -q^2\d_z\varphi^a+\frac{\g^2 \, v(z)^2}{z^2} \, \d_z \pi^a =0 \, ,
\label{Az}
\end{equation}
where $v(z) \! = \! 2X_0(z) \! = \! m_q z \! + \! \sigma z^3$.
Whereas Refs.~\cite{GR1,GR2} obtain solutions of Eq.~(\ref{eqVAdS})
(corresponding to the tower of $\rho$ states), we focus instead on
Eqs.~(\ref{AL}) and (\ref{Az}), coupled equations corresponding to the
$\pi$ and its excitations.  The dimensionful chiral symmetry-breaking
parameters $m_q$ and $\sigma$ render Eqs.~(\ref{AL}) and (\ref{Az})
[as well as Eq.~(\ref{AT})] more complicated than Eq.~(\ref{eqVAdS}),
which depends upon only one dimensionful parameter, $z_0$ or $\kappa$.

The gauge/gravity correspondence provides a method of determining the
5D gauge coupling $g_5$.  Applying the equation of motion
Eq.~(\ref{eqVAdS}) to the $F_V^2$ portion of the action
Eq.~(\ref{5DL_V_A}) leaves only the boundary term
\begin{equation}
S = -\frac{1}{2\g^2} \int\! d^4x \, \left. \frac{e^{-\Phi}}{z} V_\mu^a
\d_z V^{\mu a} \right|_{z=\epsilon} \, .
\end{equation}
The significance of this quantity becomes clear when one resolves the
vector field as $V_\mu^a (q,z) = V(q,z) \tilde V_\mu^a (q)$, where
$\tilde V_\mu^a (q)$ is the Fourier transform of the source of the
vector current $J_\mu^a \! = \! \bar q \gamma_\mu t^a q$ at the UV
boundary $z \! = \! \epsilon$, and $V(q,z)$ (the ``bulk-to-boundary
propagator'') is normalized to $V(q,\epsilon)=1$.  Due to the isospin
conservation constraint $q_\mu V^\mu \! = \! 0$, one may replace
$\tilde V_\mu^a \tilde V^{\mu \, a}$ with $\tilde V_\mu^a \tilde
V_\nu^b \Pi^{\mu \nu} \delta^{ab}$ and $\Pi^{\mu \nu} \! \equiv \!
\eta^{\mu \nu} \! - \! q^\mu q^\nu /q^2$, and then the usual quadratic
variation of the action with respect to the source $\tilde V$ produces
the vector current two-point function:
%
\begin{eqnarray}
  \int d^4 x \, e^{iqx} \<J_\mu^{a}(x)J_\nu^{b}(0)\> \!&=&\!
  \delta^{ab} \, \Pi_{\mu \nu} \, \Sigma_V (q^2) \, ,
  \label{eq:VV}\\
  \Sigma_V (q^2) \!&=&\! 
  \left.-\frac{e^{-\Phi}}{\g^2} \frac{\d_z V(q,z)}{z}
\right|_{z=\epsilon} \, ,
\label{Vqz}
\end{eqnarray}
%
from which one finds, matching to the QCD result for currents $J_\mu$
normalized~\cite{GR2} according to the prescription of~\cite{EKSS}
\begin{equation} \label{g5val}
\g^2 = \frac{12 \pi^2}{N_c} \to 4\pi^2 \,.
\end{equation}

An analogous calculation in the axial sector relates the
bulk-to-boundary propagator $A(q,z)$ to the $\pi$ decay constant
$f_\pi$:
\begin{equation}\label{fpi}
f_\pi^2 = -\frac1{\g^2}\left.\frac{\d_z
A(0,z)}{z}\right|_{z=\epsilon}.
\end{equation}

The sets of normalizable eigenstates of
Eqs.~(\ref{eqVAdS})--(\ref{Az}) form towers of hadrons of the
corresponding quantum numbers.  Since large $N_c$ is intrinsic to this
procedure, the mesons have narrow widths and the spectral
decompositions of self-energy functions such as $\Sigma_V$ are sums
over poles:
\begin{equation}
\Sigma_V (q^2) = \sum_{n=0}^\infty \frac{f_n^2}{q^2 - M_n^2} \, ,
\end{equation}
where $M_n$ are the mass eigenvalues and $f_n$ are the decay constants
of vector modes $\psi_n (z)$ normalized according to
\begin{equation}
\int dz \frac{e^{-\Phi}}{z} \, \psi_m (z) \psi_n (z) = \delta_{mn} \,
.
\end{equation}

The coupled equations of motion Eqs.~(\ref{AL})--(\ref{Az}) for the
bulk-to-boundary propagators $\d_z \varphi(q,z)$, $\d_z \pi(q,z)$ can
be combined to produce the decoupled and dimensionless Sturm-Liouville
form
\begin{equation} \label{decoupled}
\d_x \left[ \Lambda(x) \, \d_x y(x) \right] + \Lambda(x) \left[
\tilde{q}^2 - \beta(x) \right] y(x) = 0 \, ,
\end{equation}
where, taking $\mu \! \equiv 1/z_0$ or $\kappa$ in the hard- and
soft-wall cases, respectively, the dimensionless independent and
dependent variables are $x \! \equiv \! \mu z$ and $y(x) \! \equiv \!
[e^{-\Phi(x/\mu)}/x][\d_x \varphi (\mu \tilde{q}, x/\mu)]$,
respectively.  Furthermore, $\tilde{q}^2 \! \equiv q^2/\mu^2$ and
$\beta(x) \! \equiv g_5^2 \, v(\mu x)^2 / x^2$, or
\begin{equation}
\beta(x) \equiv ( \tilde{m}_q + \tilde{\sigma} x^2 )^2 \, ,
\end{equation}
where $\tilde{m}_q \! \equiv g_5 m_q / \mu$ and $\tilde{\sigma}
\! \equiv g_5 \sigma / \mu^3$ are dimensionless, and $\Lambda(x) \!
\equiv x/[\beta(x) e^{-\Phi(x/\mu)}]$.  Equation~(\ref{Az}) then
immediately gives a solution for the field $\d_z \pi (q,z)$:
\begin{equation}
\d_x \pi (\mu \tilde{q}, x/\mu) = \tilde{q}^2 \Lambda(x) \, y(x) \, .
\end{equation}

Equation (\ref{decoupled}) is therefore a second-order ordinary
differential equation in $x$ with three dimensionless parameters,
$\tilde{q}^2$, $\tilde{m}_q$, and $\tilde{\sigma}$.  It appears not to
admit a general closed-form solution in terms of well-known functions.
However, one may, as in this paper, numerically solve the equation
subject to physical constraints (fitting to $m_\rho$, $f_\pi$, and
$m_\pi$).  One may also explore limiting cases for various orderings
of the parameters; since $\tilde{\sigma}$ enters Eq.~(\ref{decoupled})
through the combination $\tilde{\sigma} x^2$, one may consider the
behavior of (\ref{decoupled}) in the limits where various ratios of
$\tilde{q}^2$, $(\tilde{\sigma} x^2)^2$, and $\tilde{m}_q^2$ are taken
to be small.

In the hard-wall case of $\Phi(x/\mu) \! = \! 1$, the analytically
soluble cases are \\
${\bf 1}) \ \ \tilde{q}^2, \tilde{m}_q^2 \gg (\tilde{\sigma} x^2)^2
\mbox{   ($z$ small or $\sigma \! = \! 0$)}:$
\begin{eqnarray}
y(x) & = & A J_0 \left( \sqrt{\tilde{q}^2 - \tilde{m}_q^2} \, x
\right) + B Y_0 \left( \sqrt{\tilde{q}^2 - \tilde{m}_q^2} \, x
\right) \, \mbox{, or} \nonumber \\
\d_z \varphi(q,z) & = & \frac{z}{z_0^2} \left[ A J_0 \left(
\sqrt{q^2 - g_5^2 m_q^2} \, z \right) + B Y_0 \left(
\sqrt{q^2 - g_5^2 m_q^2} \, z \right) \right] \, , \nonumber \\
\d_z \pi (q,z) & = & \frac{q^2 z}{g_5^2 m_q^2 z_0^2} \left[ A J_0
\left( \sqrt{q^2 - g_5^2 m_q^2} \, z \right) + B Y_0 \left(
\sqrt{q^2 - g_5^2 m_q^2} \, z \right) \right] \ .
\end{eqnarray}
For extremely small or negative $q^2$ (such that the arguments of the
square roots become negative), the Bessel functions $J_0$ and $Y_0$ of
course analytically continue to modified Bessel functions. \\
${\bf 2}) \ \ \tilde{q}^2 \gg (\tilde{\sigma} x^2)^2 \gg
\tilde{m}_q^2 \mbox{   ($q^2$ large, $z$ not extremely small)}:$
\begin{eqnarray}
y(x) & = & x^2 \left[ A J_2 (\tilde{q} x) + B Y_2 (\tilde{q} x)
\right] \, \mbox{, or} \nonumber \\
\d_z \varphi(q,z) & = & \frac{z^3}{z_0^4} [ A J_2 (qz) + B Y_2 (qz) ]
\, , \nonumber \\
\d_z \pi (q,z) & = & \frac{q^2}{g_5^2 \sigma^2 z z_0^4}
\left[ A J_2 (qz) + B Y_2 (qz) \right] \, .
\end{eqnarray}
The distinction between {\bf 1}) and {\bf 2}) arises from a
noncommutativity of limits in $\Lambda(x)$: If $\tilde{\sigma} x^2$ is
taken small first, then $\Lambda(x) \! \to \! x/\tilde{m_q}^2$, while if
$\tilde{m_q}^2$ is taken small first, then $\Lambda(x) \! \to \!
1/\tilde{\sigma}^2 x^3$. \\
${\bf 3}) \ \ (\tilde{\sigma} x^2)^2 \gg \tilde{q}^2 \gg \tilde{m}_q^2
\mbox{   ($z$ large)}:$
\begin{eqnarray}
y(x) & = & A \, {\rm Ai}^\prime \left[ \left(
\frac{\tilde{\sigma} x^3}{2} \right)^{2/3} \right] + B \,
{\rm Bi}^\prime \left[ \left( \frac{\tilde{\sigma} x^3}{2}
\right)^{2/3} \right] \mbox{, or} \nonumber \\
\d_z \varphi(q,z) & = & \frac{z}{z_0^2} \left\{ A \,
{\rm Ai}^\prime \left[ \left( \frac{g_5 \sigma z^3}{2}
\right)^{2/3} \right] + B \, {\rm Bi}^\prime \left[ \left( \frac{g_5
\sigma z^3}{2} \right)^{2/3} \right] \right\} \, , \nonumber \\
\d_z \pi (q,z) & = &  \frac{q^2}{g_5^2 \sigma^2 z^3 z_0^2} \left\{ A
\, {\rm Ai}^\prime \left[ \left( \frac{g_5 \sigma z^3}{2}
\right)^{2/3} \right] + B \, {\rm Bi}^\prime \left[ \left( \frac{g_5
\sigma z^3}{2} \right)^{2/3} \right] \right\} \, .
\end{eqnarray}
Note that all the special functions appearing in this case are
variants of Bessel functions, as seen repeatedly in previous papers
that consider solutions to Eq.~(\ref{eqVAdS}) for the hard-wall
background.

In the soft-wall case of $\Phi(x/\mu) \! = \! x^2$, the only
analytically soluble cases turn out to have large $q^2$, and the
functions are variants on Kummer functions $M(a,b,z)$, $U(a,b,z)$ (or
equivalently, confluent hypergeometric or Whittaker
functions)~\cite{AS}; similar functions have been seen for solutions
to Eq.~(\ref{eqVAdS}) for the soft-wall background~\cite{GR2}.  The
analytically soluble cases are\\
${\bf 1}) \ \ \tilde{q}^2, \tilde{m}_q^2 \gg (\tilde{\sigma} x^2)^2
\mbox{   ($z$ small or $\sigma \! = \! 0$)}:$
\begin{eqnarray}
y(x) & = & e^{-x^2} \left\{ A \, M \! \left[ 1 - {\scriptstyle{\frac 1
4}} (\tilde{q}^2 - \tilde{m}_q^2) , \, 1, \, x^2 \right] + B \, U \!
\! \left[ 1 - {\scriptstyle{\frac 1 4}} (\tilde{q}^2 - \tilde{m}_q^2)
, \, 1, \, x^2 \right] \right\} \, \mbox{, or} \nonumber \\
\d_z \varphi(q,z) & = & \kappa^2 z \left\{ A \, M \! \left[
1 - \frac{q^2 - g_5^2 m_q^2}{4\kappa^2}, \, 1, \, (\kappa z)^2 \right]
+ B \, U \! \! \left[ 1 - \frac{q^2 - g_5^2 m_q^2}{4\kappa^2}, \, 1,
\, (\kappa z)^2 \right] \right\} \, , \nonumber \\
\d_z \pi(q,z) & = & \frac{q^2\kappa^2 z}{g_5^2 m_q^2} \left\{ A \, M \!
\left[ 1 - \frac{q^2 - g_5^2 m_q^2}{4\kappa^2}, \, 1, \, (\kappa z)^2
\right] + B \, U \! \! \left[ 1 - \frac{q^2 - g_5^2 m_q^2}{4\kappa^2},
\, 1, \, (\kappa z)^2 \right] \right\} \, . \nonumber \\
\end{eqnarray}
${\bf 2}) \ \ \tilde{q}^2 \gg (\tilde{\sigma} x^2)^2 \gg
\tilde{m}_q^2 \mbox{   ($q^2$ large, $z$ not extremely small)}:$
\begin{eqnarray}
y(x) & = & x^4 e^{-x^2} \left[ A \, M \! \left( 1 - \tilde{q}^2 /4,
\, 3, \, x^2 \right) + B \, U \! \left( 1 - \tilde{q}^2 /4, \, 3,
\, x^2 \right) \right] \, \mbox{, or} \nonumber \\
\d_z \varphi(q,z) & = & \kappa^6 z^5 \left\{ A \, M \! \left[ 1 - q^2
\! /4\kappa^2, \, 3, \, (\kappa z)^2 \right] + B \, U \! \! \left[ 1 -
q^2 \! /4\kappa^2, \, 3, \, (\kappa z)^2 \right] \right\} \, ,
\nonumber \\
\d_z \pi(q,z) & = & \frac{q^2 \kappa^6 z}{g_5^2 \sigma^2} \left\{ A \,
M \! \left[ 1 - q^2 \! /4\kappa^2, \, 3, \, (\kappa z)^2 \right] + B
\, U \! \! \left[ 1 - q^2 \! /4\kappa^2, \, 3, \, (\kappa z)^2 \right]
\right\} \, .
\end{eqnarray}
Numerous well-known recursion relations, distinct in form for $M$ and
$U$, may be used to reduce the arguments of the Kummer
functions~\cite{AS}, but we opt to present expressions for which $M$
and $U$ have the same arguments.

\section{Form Factor Expressions} \label{newFF}

In the present calculations we are interested in the behavior of both
form factors and the 3-point couplings $g_{n \pi \pi}$ [or
$g_{\rho^{(n)} \pi \pi}$] between the $n$th vector state and the
lowest eigenstate of the field $\pi$.  Of course, the $n \! = \!  0$
case is the AdS/QCD version of $g_{\rho \pi \pi}$.  In order to
identify these couplings, one must expand the action
Eq.~(\ref{5DL_V_A}) to cubic order in fields.  Since Eq.~(\ref{Az})
relates the pion field to the longitudinal mode $\d^\mu \varphi$ of
$A^\mu$, one must identify not only $V\pi\pi$ terms, but also $V \!
AA$ and $V \! A\pi$.  Schematically, $DX \sim \d \pi + \pi \d \pi +
O(\pi^3) + V \pi + O(V \pi^3) + A + A \pi + O(A \pi^2)$, $F_V \! \sim
\! \d V \! + \! VV \!  + \! AA$, and $F_A \! \sim \! \d A \! + \!
VA$.  $X$, the only field carrying SU(2) fundamental representation
indices, must appear at least in pairs, while the $\d A$ term of $F_A$
contains no longitudinal piece: $\d^\mu (\d^\nu \varphi) - \d^\nu
(\d^\mu \varphi) = 0$.  The relevant terms then arise from the cross
terms $(\d \pi)(V \pi)$ and $(V \pi)(A)$ of $|DX|^2$ and $(\d V)(AA)$
of $F_V^2$.  One obtains the $V\pi\pi$ terms
\begin{align}\label{Vpipi_Action}
S_{\rm AdS}^{V\pi\pi}  &=
 \epsilon_{abc}\int d^4 x \int dz \, e^{-\Phi} \,
 \left[  \frac{1}{g_5^2\,z}
  \left(\partial_z\partial^\mu\varphi^a\right) V_{\mu}^b
  \left(\partial_z\varphi^c\right) \right. \nonumber \\
 &\left. \quad +\frac{v(z)^2}{z^3}
  \left(\partial^{\mu}\pi^a- \partial^{\mu}\varphi^a\right)
  V_{\mu}^b \left(\pi^c-\varphi^c\right)
  \right] \, .
\end{align}
where the integration ranges over $[0,z_0]$ in the hard-wall case and
$[0,\infty)$ in the soft-wall case~\cite{thanksjosh}.
Reference~\cite{EKSS} uses this action (not including the dilaton
coupling) to obtain the $V \pi \pi$ couplings [Eq.~(\ref{gnpipi})
below], with the caveat that terms cubic in $F_{V,A}$ have not been
included.  In fact, we now show that no such terms contribute to the
$V \pi \pi$ coupling.

The field strength tensors $F_{V,A}^{MN}$ are antisymmetric and of
opposite parities.  Due to antisymmetry, only one $F_V^3$ term
($\equiv F_V{}^L{}_M F_V{}^M{}_N F_V{}^N{}_L$) and one $F_V F_A^2$
term occurs, while the terms with an odd number of $F_A$'s are
pseudoscalars.  It is tempting to remove the parity distinction by
forming the dual ${\cal F}_A$ of $F_A$, but in 5D this object is a
rank-3 tensor:
\begin{equation}
{\cal F}_{JKL} \! \equiv \! \frac 1 6
\epsilon^{\vphantom\dagger}_{JKLMN} F^{MN} \, .
\end{equation}
Pairs of Levi-Civita tensors may always be converted into metric
tensors using (5D versions of) the usual identities, so the only
additional terms one might consider have a single ${\cal F}_A$.  But
such terms cannot form scalars because they have an odd total number
of Lorentz indices.  Thus, only $F_V^3$ and $F^{\vphantom{2}}_V F_A^2$
need be considered.

However, $F_V^3$ terms that are linear in $V$ contain at least 5
$\pi$'s, while the surviving terms of $F_A$ contain at least one $V$,
and therefore all terms in $F^{\vphantom{2}}_V F_A^2$ are at least
quadratic in $V$.  It follows that none of the $F^3$ terms contribute
to the $V \pi \pi$ coupling.

Returning to the action Eq.~(\ref{Vpipi_Action}), a naive variation
gives the 3-point correlator:
\begin{align}
 \langle J_\pi^a(p_1)J_V^{\mu,b}(q) J_\pi^c(-p_2)\rangle =
 \epsilon^{abc} F(p_1^2,p_2^2,q^2)\left(p_1+p_2\right)^\mu i (2\pi)^4
\delta^{(4)} (p_1 - p_2 + q) \ .
\end{align}
Again recalling the narrowness of resonances, one may express the
dynamical factor $F(p_1^2,p_2^2,q^2)$ in terms of transition form
factors:
\begin{align}
\label{Fsuv} F(p_1^2,p_2^2,q^2) = \sum_{n,k = 1}^{\infty}
 \frac{f_{n} f_{k}  F_{nk} (q^2) }{\left(p_{1}^2 -
 M^2_{n}\right)\left(p_{2}^2 - M^2_{k}\right)}  \ ,
\end{align}
where $F_{nk}(q^2)$ correspond to form factors for $n \! \to \! k$
transitions.  The pion form factor $F_\pi (q^2)$ is then obtained as
\begin{equation}
\label{ff} F_\pi(q^2) \equiv F_{11}(q^2) = \int dz \, e^{-\Phi} \,
\frac{V(q,z)}{f_\pi^2} \left\{ \frac{1}{g_5^2 z}
[\partial_z\varphi(z)]^2 + \frac{v(z)^2}{z^3} \left[\pi(z) -
\varphi(z)\right]^2 \right\} \ ,
\end{equation}
an expression whose origin may be recognized in
Eq.~(\ref{Vpipi_Action}).  It is directly derived from that equation
by factoring the 5D fields into products of the (dimensionless) 4D
pion fields $\pi^a(q)$ and the bulk-to-boundary propagators $\pi(z)$
and $\varphi(z)$.  In order for the $\pi^a(q)$ kinetic energy term to
receive the standard canonical normalization $\frac 1 2 \partial_\mu
\pi^a \partial^\mu \pi^a$, the integral in Eq.~(\ref{Vpipi_Action})
[at $q^2 \! = \! 0$, where $V(q,z) \! = \!  1$] must equal unity,
which fixes the normalization of Eq.~(\ref{Vpipi_Action}).  Note that
the explicit $f_\pi^2$ factor in Eq.~(\ref{Vpipi_Action}) is part of
the normalization and does not change the shape of $F_\pi (Q^2)$.  The
pion is the ground-state solution to Eqs.~(\ref{AL})--(\ref{Az})
subject to the constraints at the large-$z$ termini described in
Sec.~\ref{formalism}.  The integral in Eq.~(\ref{ff}), setting $V(q,z)
\! = \! 1$, is normalized to unity, giving a canonically-normalized
kinetic energy term for the 4D pion field.  Taking a spacelike
momentum transfer $q^2 \! \equiv \! -Q^2$ for the vector source $V$
that solves Eq.~(\ref{eqVAdS}) gives~\cite{GR1}
\begin{equation} \label{Vhard}
V(q,z) \equiv {\cal J}(Q,z) = {Qz} \left  [ K_1(Qz) + I_1(Qz)
\frac{K_0(Qz_0)}{I_0(Qz_0)} \right  ] \, ,
\end{equation}
for the hard-wall case, while the corresponding expression for the
soft-wall case is~\cite{GR2,BdT}
\begin{equation} \label{Vsoft}
V(q,z) \equiv {\cal J}(Q,z) = \Gamma (1 + Q^2 \! /4\kappa^2) \, U[
Q^2 \! /4\kappa^2, 0, (\kappa z)^2 ] \, .
\end{equation}
Both of these solutions satisfy the boundary conditions $V(q,\epsilon)
\! = \! 1$, $V(0,z) \! = \! 1$, as well as $\d_z V(q,z) \! = \! 0$ for
$z \! = \! z_0$ in the hard-wall case.  In addition, for large $z$
Eq.~(\ref{Vsoft}) falls as $(z^2)^{-Q^2/4}$.  Expressing instead
$V(q,z)$ for timelike momentum transfers gives
\begin{equation}
 \label{vector_expand}
 V(q,z) = -g_5 \sum_{n = 1}^{\infty} \frac{f_{n} \psi_n( z)}
 { q^2 - M^2_{n} } \, .
\end{equation}
Substituting this expression into Eq.~(\ref{ff}), one can represent
the timelike pion form factor as a sum over vector meson poles:
\begin{equation}
 \label{pion_timelike}
 F_\pi(q^2) = -\sum_{n = 1}^{\infty} \frac{f_{n} g_{n\pi\pi}}
 {q^2 - M^2_{n}} \ ,
\end{equation}
where $g_{n\pi\pi}$ is given by
\begin{equation} \label{gnpipi}
g_{n \pi \pi} = \frac{g_5}{f_\pi^2} \int \! dz \, \psi_n (z) e^{-\Phi}
\left\{ \frac{1}{g_5^2 z} [\d_z \varphi (z)]^2 + \frac{v(z)^2}{z^3}
\left[ \pi (z) - \varphi(z) \right]^2 \right\} \, \ .
\end{equation}
Together, Eqs.~(\ref{ff}) with Eqs.~(\ref{Vhard}) or (\ref{Vsoft}) and
Eqs.~(\ref{pion_timelike})--(\ref{gnpipi}) provide a complete
expression for the pion form factor in all kinematic regions.

\section{Results} \label{results}
In this section we present numerical predictions for QCD observables
in both the hard- and soft-wall models.
We perform the fit for the three hard-wall parameters
$z_0$ ($z_m$ in \cite{EKSS}), $m_q$, and $\sigma$ to the three
observables $m_\rho$, $m_\pi$, and $f_\pi$.  To the same observables
we also fit the three parameters $\kappa$, $m_q$, and $\sigma$ of the
soft-wall model.  In the hard-wall case the $\rho$ wave functions
[eigenfunctions of Eq.~(\ref{eqVAdS})] are Bessel functions, with
masses determined by zeroes of $J_0(q z_0)$; hence, $m_\rho \! = \!
\gamma_{0,1}/z_0 \! = \! 775.5$~MeV, where $\gamma_{0,1} \! = \! 2.405$
fixes $z_0 \!  = \!  1/(322\ {\rm MeV})$.  One may then fit $m_q$ and
$\sigma$ to the experimental values of $m_\pi$ and $f_\pi$ [which are
constrained by the Gell-Mann--Oakes--Renner (GMOR) relation $m_\pi^2
f_\pi^2 \! = \! 2m_q \sigma$], yielding $m_q \! = \!  2.30$ MeV and
$\sigma \!  = \! (326 \ {\rm MeV})^3$ for the hard-wall model.  In the
soft-wall model the vector mass eigenvalues grow linearly in $n$
[$m_{\rho,n}^2 \! = 4(n \! + \!  1)\kappa^2$]; hence, the $\rho$ ($n
\! = \! 0$) mass fixes $\kappa \!  = \! m_\rho/2 \! = \!  389 \ {\rm
MeV}$.  Note in particular that $m_\rho$ is fixed entirely by the
value of $\mu \! = \! 1/z_0$ or $\kappa$; $m_q$ and $\sigma$ can then
be fit to the experimental values of $m_\pi$ and $f_\pi$, yielding for
the soft-wall model $m_q \! = \!  1.45$~MeV and $\sigma \! = \! (368\
{\rm MeV})^3$.
\begin{figure}[t]
\epsfxsize 5.0 in \epsfbox{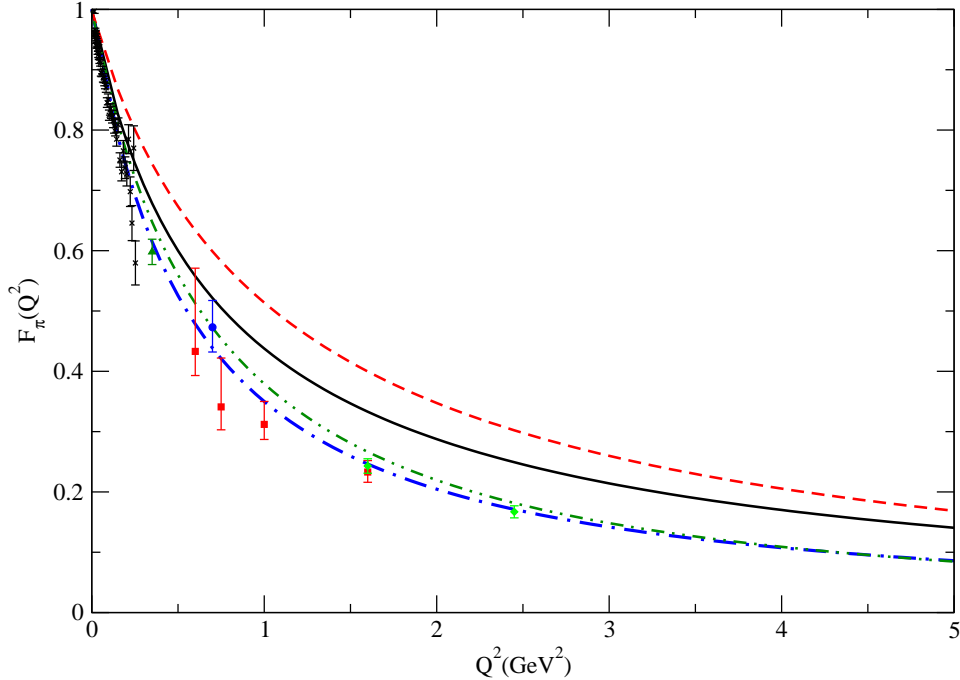}
\caption{Spacelike scaling behavior of $F_\pi(Q^2)$ as a function of
$Q^2 \! = \! -q^2$.  The continuous line is the prediction of the
original hard-wall model.  The dashed line is the prediction of the
original soft-wall model with $\kappa \! = \! m_\rho/2$.  The dash-dot
line is the hard-wall model with $\sigma \! = \! (254 \ {\rm MeV})^3$,
and the dash-double-dot line is the soft-wall model with $\sigma \!  =
\! (262 \ {\rm MeV})^3$.  The
crosses are from a data compilation from CERN~\cite{Ame84},
the
circles are from DESY, reanalyzed by Tadevosyan {\it et
al.}~\cite{Bra77,Tadevosyan:2007yd}, the
triangle is data from DESY~\cite{Ack78}, and the
boxes~\cite{Tadevosyan:2007yd} and
diamonds~\cite{Horn:2006tm} are from Jefferson Lab.  Older data in the
range 3--10~GeV$^2$~\cite{bebek} exist but have large uncertainties
and are not plotted here.}
\label{fig:formfactor}
\end{figure}
%
\begin{figure}[t]
\epsfxsize 5.0 in \epsfbox{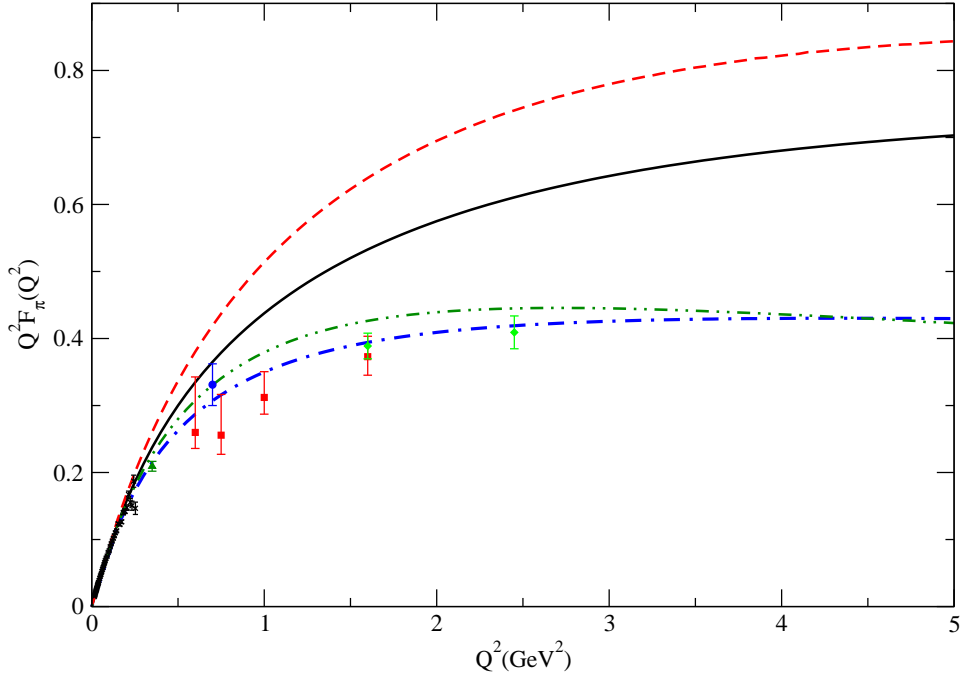}
\caption{Spacelike scaling behavior of $Q^2 F_\pi (Q^2)$ as a function of
$Q^2 \! = \! -q^2$.  The symbols are the same as in
Fig.~\ref{fig:formfactor}.
}
\label{fig:Q2formfactor}
\end{figure}

Predictions for other QCD observables in both models are collected in
Table~\ref{Table}.
Already one sees that the soft-wall model predicts
$f_\rho$ to be much smaller than the experimental data, while the
hard-wall model predicts a much closer value.  Indeed, as noted in
Ref.~\cite{GR2}, the soft-wall model predicts the ratio
$m_\rho^2/f_\rho$ to be exactly $2\sqrt{2}\pi \! = \! 8.89$, which
differs dramatically from the experimental value $5.02 \! \pm \! 0.04$
obtained from Table~\ref{Table}; in comparison, the hard-wall model
predicts this ratio to be $\sqrt{2} \pi \gamma_{0,1} J_1
(\gamma_{0,1}) \!  = \! 5.55$.

In Figs.~\ref{fig:formfactor} and \ref{fig:Q2formfactor} we plot the
electromagnetic pion form factor $F_\pi (Q^2)$ in the spacelike region
as obtained from both models.  While results for both models are too
shallow for all $Q^2$, the hard-wall (solid line) prediction tends to
lie closer to the experimental data than the soft-wall (dashed line)
prediction.  The hard-wall fit to $F_\pi (Q^2)$ can be improved for
smaller $Q^2$ by reducing $1/z_0$ to about 255~MeV, at the expense of
the vector meson parameters: $m_\rho \! = \! 613$~MeV and
$f_\rho^{1/2} \! = \! 260$~MeV.  In the soft-wall model, if one
changes the value $\kappa$ from 388~MeV to 516~MeV in order to fit
$f_\rho$ rather than $m_\rho$ (which becomes $1032$~MeV), the result
for $F_\pi (Q^2)$ lies even further from the experimental data.  On
the other hand, if one lowers the value for $\kappa$, which results in
worse predictions for both $f_\rho$ and $m_\rho$, then the fit to
$F_\pi (Q^2)$ lies closer to the experimental data; for example,
$\kappa \! = \! 300$~MeV (not plotted) gives the physical $m_\pi$ and
$f_\pi$ values by taking $\sigma \!  = \!  (364 \ {\rm MeV})^3$ and
$m_q \! = \! 1.70$~MeV, but then $m_\rho \! = \!  600$~MeV and
$f_\rho^{1/2} \! = \! 201$~MeV.  Lowering $\kappa$ to 255~MeV turns
out to match $F_\pi(Q^2)$ data somewhat better, but then $m_\rho \!  =
\!  510$~MeV and $f_\rho^{1/2} \! = \!  171$~MeV.

As noted above, once the vector parameters are determined by the value
of $\mu \! = \!  1/z_0$ or $\kappa$, the pion sector of the models
determines best fit values for $\sigma$ and $m_q$.  One finds using
Eq.~(\ref{fpi}) that $f_\pi$ depends mostly on $\sigma$, and $m_\pi$
is then fixed by choosing $m_q$ to satisfy the GMOR relation.  Since,
as is apparent in Figs.~\ref{fig:formfactor} and
\ref{fig:Q2formfactor}, the $F_\pi (Q^2)$ prediction from neither
model is particularly good as $Q^2$ increases, we consider the effect
upon $F_\pi (Q^2)$ of varying $\sigma$ and $m_q$ in both models.
Empirically, both of the observables scale very close to the square
root of the parameters ($m_\pi^2 \! \propto \! m_q$ and $f_\pi^2 \!
\propto \! \sigma$), but as one might expect, $F_\pi (Q^2)$ depends
much more strongly upon $\sigma$.  Hence, if one allows $\sigma$ to
float to fit the data for $F_\pi (Q^2)$, the precise fit to $f_\pi$ is
spoiled.  In particular, lowering the value for $\sigma$ to $(254 \
{\rm MeV})^3$ in the hard-wall model but leaving $1/z_0 \! = \!
322$~MeV gives a much better fit to $F_\pi (Q^2)$ (dash-dot line in
Figs.~\ref{fig:formfactor} and \ref{fig:Q2formfactor}), but at the
price of lowering the prediction of $f_\pi$ to $64.2$~MeV.  In the
soft-wall model, lowering the value for $\sigma$ to $(262 \ {\rm
MeV})^3$ and leaving $\kappa \! = \!  389$~MeV gives a much better fit
to $F_\pi (Q^2)$ (dash-double-dot line in Figs.~\ref{fig:formfactor}
and \ref{fig:Q2formfactor}), but at the price of lowering the
prediction of $f_\pi$ to $52.2$~MeV.

Using these form factor results for very low $Q^2$, one can extract the
pion charge radius $\langle r_\pi^2\rangle \! \equiv \! - 6 dF_\pi
(Q^2)/dQ^2 |_{Q^2 = 0}$.  The experimental value $\langle r_\pi^2\rangle
\! = \! [0.672(8) \ {\rm fm}]^2$~\cite{PDG} lies closer to the original
hard-wall [$\langle r_\pi^2\rangle \! = \!  (0.576 \ {\rm fm})^2]$ than
the soft-wall [$\langle r_\pi^2\rangle \! = \! (0.494 \ {\rm fm})^2$]
results, as a glance at Fig.~\ref{fig:formfactor} suggests.  As remarked
above, the hard-wall model fits the data better with $\sigma \! = \! (254
\ {\rm MeV})^3$, from which one finds $\langle r_\pi^2\rangle \! = \!
(0.645 \ {\rm fm})^2$, while setting $\sigma \! = \! (262 \ {\rm MeV})^3$
for the soft-wall model gives $\langle r_\pi^2\rangle \! = \!  (0.600
\ {\rm fm})^2$.

That the soft-wall model gives a shallower prediction for
$F_\pi(Q^2)$, and hence a smaller value for $\langle r_\pi^2\rangle$
compared to the hard-wall prediction, is quite easy to explain
numerically.  First note that $V(Q,z)$ in Eqs.~(\ref{Vhard}) and
(\ref{Vsoft}) have quite similar $z$ behaviors, except that $V(Q,z)$
for the hard wall is cut off at $z \! = \! z_0$.  This fact alone
allows for a greater contribution to the integral in Eq.~(\ref{ff}) in
the soft-wall case.  Moreover, the soft wall allows for more
penetration of the $\pi(z)$ and $\varphi(z)$ fields into the bulk, as
verified by our numerical simulations.  So together they contribute
more to the integration in Eq.~(\ref{ff}) and give a higher value of
$F_\pi (Q^2)$ at any particular value of $Q^2$ than for the hard-wall
model.

Finally, we comment upon vector meson dominance and $F_\pi (q^2)$ in
the timelike region.  In the hard-wall case we find that $F_\pi (0)$
is essentially saturated by the first three $\rho$ meson poles.
Explicitly, from the values $F_n \! = \! \gamma_{0,n} / [z_0^2
\sqrt{2} \pi |J_1 (\gamma_{0,n})|]$ and $M_n \! = \!
\gamma_{0,n}/z_0$, and using Eq.~(\ref{gnpipi}) to compute
%
$g_{1\pi\pi} \! = \! 2.3616$, $g_{2\pi\pi} \! = \! -0.8968$, we
find the first few contributions to $F_\pi (0)$ of $0.8079 \! + \!
0.2830 \!  - \!  0.0859 \! = \! 1.0050$.  However, the convergence is
much slower in the soft-wall model.
In the soft-wall case one has $F_n \! = \! \kappa^2 (\sqrt{2}/\pi)
(n+1)^{1/2}$ and $M_n = 2\kappa (n+1)^{1/2}$, and computes from
Eq.~(\ref{gnpipi}) the couplings
%
%
$g_{1\pi\pi} \! = \! 3.3882$, $g_{2\pi\pi} \! = \! 2.9157$, and
$g_{3\pi\pi} \! = \! 2.2946$.
With these values, we find the contributions to $F_\pi (0)$ of $0.3751 \! 
+ \! 0.2696 \! + \! 0.1894 \! + \! 0.1291 \!
= \! 0.9632$.
%
%
The next five terms are also positive and bring the sum of pole
contributions to $1.138$ before turning negative, a pattern that
persists for several terms and brings $F_\pi(0)$ back close to 1\@.
In contrast, results obtained in Refs.~\cite{GR1,GR2} for the form
factor $F_\rho (Q^2)$ show very different behavior, requiring fewer
resonances for saturation in the soft-wall case.
\begin{table}[t]\vspace{-6pt}
\caption[]{Hard- and soft-wall model predictions for QCD observables,
the three model parameters in each case fit to $m_\pi$, $f_\pi$, and
$m_\rho$ (indicated by asterisks); all values except $g_{\rho \pi
\pi}$ are in MeV.}
\begin{ruledtabular}
\begin{tabular}{cccc}  
Observable         & Experiment                           & Hard-wall     &Soft-wall\\
\hline
$m_\pi$            & 139.6$\pm0.0004$ \cite{PDG}          & 139.6$^*$     & 139.6$^*$ \\
$m_\rho$           & 775.5$\pm 0.4$   \cite{PDG}          & 775.3$^*$     & 777.4$^*$   \\
$m_{a_1}$          & 1230$\pm40$      \cite{PDG}          & 1358          & 1601  \\
$f_\pi$            & 92.4$\pm0.35$    \cite{PDG}          & 92.1$^*$      & 87.0$^*$  \\
$f_\rho^{\,1/2}$   & 346.2$\pm1.4$    \cite{Donoghue}     & 329           & 261   \\
$f_{a_1}^{\,1/2}$  & 433$\pm13$     \cite{SS,Isgur:1988vm}& 463           & 558   \\
$g_{\rho\pi\pi}$   & 6.03$\pm0.07$    \cite{PDG}          & 4.48          & 3.33  \\
\end{tabular}
\end{ruledtabular}
\label{Table}
\end{table}

\section{Discussion and Conclusions} \label{concl}
In this paper we have considered the problem of pion dynamical
properties in holographic QCD, specifically in the context of models
for which chiral symmetry breaking, both spontaneous and explicit, is
incorporated.  In this way it differs from the recent work in
Ref.~\cite{BdT}, which considers similar problems but contains no
parameters analogous to $\sigma$ and $m_q$, and therefore uses a much
simpler form for the pion bulk-to-boundary propagator.  Nevertheless,
the basic result that $F_\pi (Q^2)$ is steeper in the hard-wall than
the soft-wall model is common to both calculations.

We also present explicit expressions for the pion bulk-to-boundary
propagator $\d_z \pi (q,z)$ in all regimes where closed-form analytic
solutions are possible.  While not directly used in our numerical
analysis of $F_\pi (Q^2)$, these results are useful for processes
involving the tower of $\pi^{(n)}$ pseudoscalars.

In our numerical studies of $F_\pi (Q^2)$ we find that the naive
hard-wall model appears to be more satisfactory than the soft-wall
model over all regions of $Q^2$, although the both models can be
quantitatively improved by tweaking their parameters (at the expense of
fits to other observables).  Even before the $F_\pi (Q^2)$ data is
included, one can argue in favor or against either model based upon
certain features; the soft-wall model has appropriate linear Regge
trajectories but poor agreement for the $m_\rho^2/f_\rho$ ratio, while
the hard-wall model is simpler and has the opposite behavior.  The
inclusion of the data for $F_\pi (Q^2)$ suggests an improved model
with a semi-hard wall, such as provided by a quasi-``Saxon-Woods''
background:
\begin{equation}
e^{-\Phi(z)} = \frac{e^{\lambda^2 z_0^2} - 1}{e^{\lambda^2 z_0^2} +
e^{\lambda^2 z^2} - 2} \, ,
\end{equation}
which has a drop-off at $z \! = \! z_0$ but falls off as
$e^{-\lambda^2 z^2}$ for large $z$, thus capturing features of both
models and allowing for better predictions.  The challenge, as always,
is to predict the most observables with the fewest parameters.

\vspace{-3.0ex}
\section*{Acknowledgments}
\vspace{-2.0ex}
We thank Andrei Belitsky and Josh Erlich for valuable discussions.
This work was supported by the NSF under Grant No.\ PHY-0456520.

\end{document}